# GALLIUM ARSENIDE PREPARATION AND QE LIFETIME STUDIES USING THE ALICE PHOTOCATHODE PREPARATION FACILITY

N. Chanlek, R.M. Jones, Cockcroft Institute / University of Manchester, Manchester, M13 9PL, UK
J.D. Herbert, L.B. Jones, K.J. Middleman and B. L. Militsyn,
Cockcroft Institute / ASTeC, STFC Daresbury Laboratory, Warrington, WA4 4AD, UK

## Abstract

In recent years, Gallium Arsenide (GaAs) type photocathodes have become widely used as electron sources in modern Energy Recovery Linac based light sources such as the Accelerators and Lasers in Combined Experiments (ALICE) at Daresbury Laboratory and as polarised electron source for the proposed International Linear Collider (ILC). Once activated to a Low Electron Affinity (LEA) state and illuminated by a laser, these materials can be used as a high-brightness source of both polarised and un-polarised electrons.

This paper presents an effective multi-stage preparation procedure including heat cleaning, atomic hydrogen cleaning and the activation process for a GaAs photocathode. The stability of quantum efficiency (QE) and lifetime of activated to LEA state GaAs photocathode have been studied in the ALICE load-lock photocathode preparation facility which has a base pressure in the order of $10^{-11}$ mbar. These studies are supported by further experimental evidence from surface science techniques such as X-ray Photoelectron Spectroscopy (XPS) to demonstrate the processes at the atomic level.

## INTRODUCTION

Current and proposed linear colliders, energy recovery linacs and new generation light sources require high quality electron sources. In particular, an electron source with high current, low emittance and good operational lifetime is desirable. The photocathode selected for the ALICE ERL is Cs:GaAs, driven by a pulsed laser at 532 nm. Photocathodes activated to NEA *in-situ* generally achieve at QE above 3 %, allowing the generation of electron bunches with charges (Q) well above 100 pC [1].

An upgrade to the ALICE photoinjector electron gun is currently underway. This will provide a photocathode preparation facility (PFF) with a load-lock interface (see Fig. 4), permitting the replacement of a cathode without breaking the vacuum in the gun. This facility also improves performance of the GaAs photocathode as the preparation procedure is more controllable. With a custom photocathode from the Institute of Semiconductor Physics, the system has demonstrated a quantum yield of 15 % at a wavelength of 635 nm [2, 3].

In spite of this success, the preparation processes for the NEA GaAs photocathode need to be optimised further. Optimal performance can be achieved through better understanding of the individual preparation processes and material properties such as surface chemistry and bulk crystallinity. To this end, we have also developed a photocathode preparation and surface analysis system (Fig. 1) which permits NEA GaAs photocathode preparation in conjunction with the application of several surface science techniques within the same vacuum system. This system provides information on surface properties, allowing optimization of cathode processing and supporting work in the photocathode preparation facility.

This paper presents the current progress and results to-date in the surface analysis system and the ALICE photocathode preparation system. An effective heat-cleaning procedure is presented with a study of the quantum efficiency stability and lifetime of the NEA GaAs photocathode.

## SURFACE ANALYSIS SYSTEM

As shown in Fig. 2, the system consists of a *load-lock chamber* where a photocathode sample is introduced into the vacuum. The sample is then transferred into the *cleaning chamber* where it can be subjected to heat cleaning and/or atomic hydrogen cleaning. The activation is carried out by applying alternate caesium and oxygen layers on clean GaAs surface in the *preparation chamber* at a base pressure better than $10^{-10}$ mbar. The *analysis chamber* supports several surface analysis techniques to characterise and study the properties of the photocathode material including X-ray Photoelectron Spectroscopy (UPS/XPS) and Low Energy Electron Diffraction (LEED).

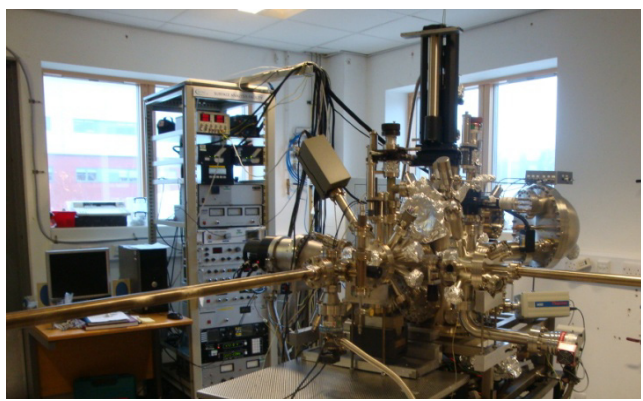

Figure 1: The surface analysis system.

### Photocathode Preparation

Our preliminary studies focused on bulk VGF (Vertical Gradient Freeze) GaAs samples with p-type Zn doping, the same photocathode specification used in the ALICE photoinjector. Before introduction to the preparation system, the samples were rinsed in high-purity iso-propanol. Once UHV conditions were established, the XPS spectra of the GaAs surface were taken using Al Kα (1486.6 eV) radiation. The typical contaminants for GaAs which have a detrimental effect on photocathode performance were observed, such as oxygen, carbon and chlorine. A clean GaAs surface is essential for successful activation because the Cs/O layers are extremely sensitive to contamination. The sample can be cleaned by heating to a temperature which is sufficient to desorb contaminants, but not so high that the stoichiometry of the surface is damaged. To optimise this heating temperature, the samples were heated to five different temperatures; 450, 500, 550, 600 and 625 °C for 60 minutes and the XPS spectra were taken *before* and *after* the heat-cleaning process. The removal by heat-cleaning of oxides which are the main surface contaminants is shown in Fig. 2.

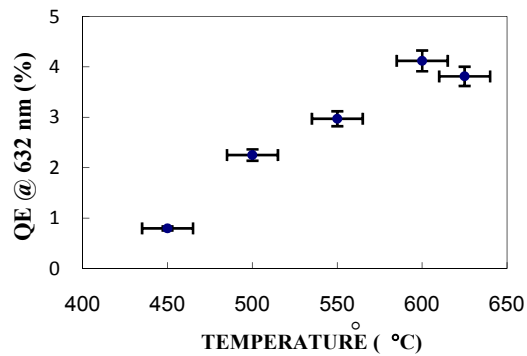

Figure 3: QE as a function of heat-cleaning temperature.

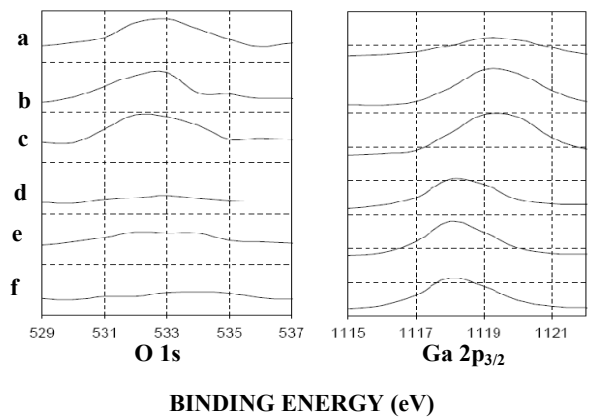

Figure 2: XPS spectra of the Ga $2p_{3/2}$ and O 1s for:
**(a)** the initial *dirty* sample; **(b)** after heating to 450 °C;
**(c)** after heating to 500 °C; **(d)** after heating to 550 °C;
**(e)** after heating to 600 °C and **(f)** after heating to 625 °C.

The presence of oxide contaminants can be identified from the O1s peak, and the shift of Ga $2p_{3/2}$ (1117 eV) peak toward higher energies. The results show that the surface oxides are desorbed from the GaAs surface under heat-cleaning, and are completely removed after heating to a temperature higher than 550 °C for 60 min.

Activation to the NEA state was made after heating the sample to each of the different temperatures. Caesium and oxygen were deposited onto the sample using the standard "*Yo-Yo*" method [4]. QE was measured using a picoammeter and HeNe laser at 632.8 nm, and the results are shown in Fig. 3. A QE of 4.12 % was obtained as a result of heating the sample at 600 °C, however, the QE dropped to 3.81 % after heating to 625 °C. The reduction in QE might be the result of excessive heating temperatures which damages the GaAs surface through the preferential desorption of arsenic from the lattice.

For this reason, we have recently installed atomic hydrogen cleaning as an effective low-temperature cleaning procedure for the preparation of NEA GaAs surfaces [5], and the thermal calibration of this modified process is currently in progress.

## ALICE PHOTOCATHODE PREPARATION

Fig. 4 shows the photocathode preparation facility which has been commissioned at the Daresbury Laboratory. The system consists of three chambers separated by manual gate valves. The three chambers from left to right are the *loading chamber*, the *cleaning chamber* and the *preparation chamber*. Photocathodes are first degreased in high-purity iso-propanol and are then introduced via a transport vessel to the loading chamber where they are introduced to the UHV system. The sample can be heated and exposed to atomic hydrogen in the cleaning chamber. Clean photo-cathodes are then transferred to the activation chamber which has a base pressure in the order of $10^{-11}$ mbar. Photocathodes undergo a final heat-cleaning, followed by activation to NEA with (Cs,O) or (Cs,NF$_3$).

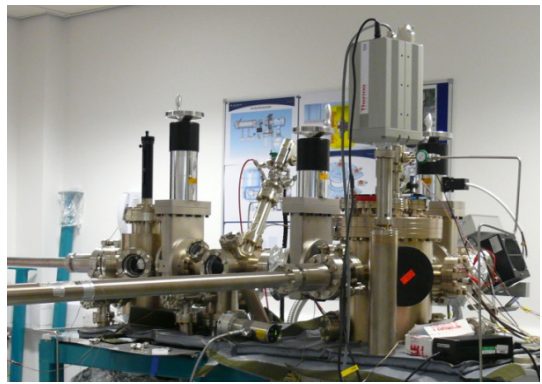

Figure 4: The ALICE photocathode preparation system.

We have defined the following preparation procedure for use in this system: the sample is heated to 550° C for 30 minutes in the preparation chamber. After cooling, caesium and oxygen are applied onto the sample surface using the standard "*Yo-Yo*" method, with a typical activation curve shown in Fig. 5. We routinely obtain QE as high as 8.5 % at 635 nm from a bulk VGF GaAs sample, identical to those currently used in the ALICE photoinjector. This level of QE is *higher* than that obtained in both the surface analysis system and following in-situ activation in the ALICE photoinjector.

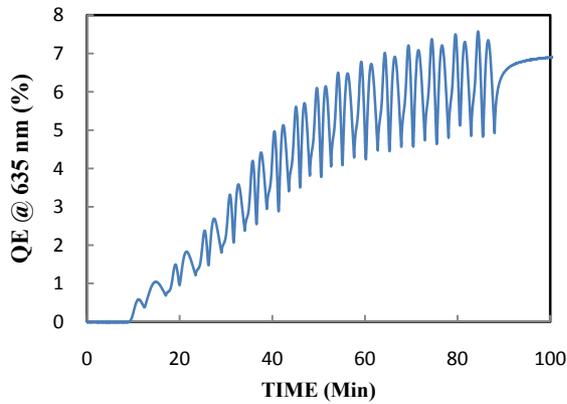

Figure 5 – Typical activation "*Yo-Yo*" curve.

*Stability of Quantum Efficiency (QE)*

One of the important parameters of the photocathode concerning their practical use is the operational lifetime because the QE of the photocathode is degraded by the internal stability of the activation layer and by the presence of oxidizing gases in the residual atmosphere of the vacuum system. In the ALICE photocathode preparation chamber, the base pressure is typically $1.5 \times 10^{-11}$ mbar, yielding $1/e$ lifetimes of 400 hrs under illumination by a CW laser delivering 0.45 mW modulated at 3 kHz, as shown in Fig. 6.

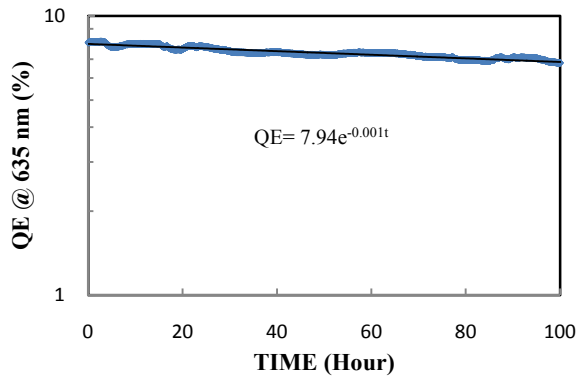

Figure 6: QE Lifetime plot for an activated photocathode.

We have studied the effect of residual gasses on the stability of the GaAs photocathode by exposing activated cathodes to several gas species in a chamber base pressure of about $1.5 \times 10^{-11}$ mbar, specifically $O_2$, CO, $CO_2$ and $CH_4$. The influence of these gasses on the QE stability is shown in Fig. 7. The data show that $CH_4$ has no effect on the QE, whilst $O_2$, CO, $CO_2$ have a significant effect. Oxygen at a pressure of $1.5 \times 10^{-10}$ mbar reduces the QE from 6 % to 1 % in 3.4 minutes.

## SUMMARY

The GaAs photocathode preparation process has been studied and optimised using a single vacuum system which allows preparation of the photocathode followed by characterisation with XPS and LEED. The heat-cleaning procedure for the GaAs photocathode was also studied. It was confirmed that oxide removal from the GaAs surface can be achieved by heating to a temperature higher than 550 °C for 60 minutes. This cleaning procedure has been applied in conjunction with our activation procedure to prepare photocathodes for ALICE with QEs around 8.5 % at 635 nm. The stability of activated GaAs photocathodes exposed to different gas species (specifically: $O_2$, $CO_2$, CO and $CH_4$) has been studied. It was found that $O_2$, $CO_2$, CO can poison the photocathode very quickly while $CH_4$ has no apparent effect on the QE of the photocathode.

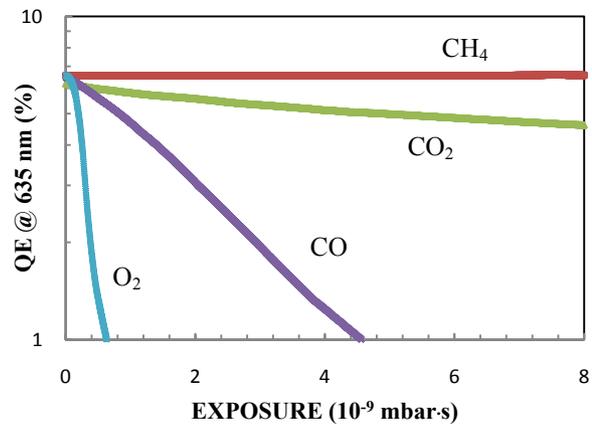

Figure 7: Decrease of QE of the photocathode due to exposure to different gasses.